\pdfoutput=1

\documentclass
[showpacs,preprintnumbers,superscriptaddress,byrevtex,onecolumn,balancelastpage,aps]{revtex4}%
\usepackage{amsmath}
\usepackage{dcolumn}
\usepackage{bm}
\usepackage{graphicx}
\usepackage{amsfonts}
\usepackage{amssymb}%
\providecommand{\U}[1]{\protect\rule{.1in}{.1in}}
\ifx\pdfoutput\relax\let\pdfoutput=\undefined\fi
\newcount\msipdfoutput
\ifx\pdfoutput\undefined\else
\ifcase\pdfoutput\else
\ifx\paperwidth\undefined\else
\ifdim\paperheight=0pt\relax\else\pdfpageheight\paperheight\fi
\ifdim\paperwidth=0pt\relax\else\pdfpagewidth\paperwidth\fi
\fi\fi\fi
\begin{document}
\preprint{ }
\title{Acoustic cloaking and imaging using complementary media}
\author{Jin Han}
\affiliation{Physics Department, Tongji University, Shanghai 200092, China}
\author{Yuancheng Fan}
\affiliation{Physics Department, Tongji University, Shanghai 200092, China}
\author{Hongqiang Li}
\email{hqlee@tongji.edu.cn}
\affiliation{Physics Department, Tongji University, Shanghai 200092, China}
\author{Zhanshan Wang}
\affiliation{Physics Department, Tongji University, Shanghai 200092, China}

\begin{abstract}
Base on the concept of complementary media, a scheme to achieve cloak an
object outside the shell is proposed by Yun lai et al [PRL 102, 093901].We
generalize the concept in acoustic waves and demonstrate numerically how to
achieve the acoustic cloaking and imaging with complementary media.

\end{abstract}

\maketitle

Making use of form-invariant coordinate transformation of Maxwell's equation,
J. B. Pendry and Leonhardt independently proposed the invisibility cloak and
the concept of transformation media for electromagnetic (EM) waves \cite{1,2}.
Transformation media transport electromagnetic (EM) waves without reflection,
giving rise to exotic EM devices such as wave concentrator\cite{3},
cylindrical superlens\cite{4}, etc., besides invisibility
cloak\cite{5,6,7,8,9}. From the view of transformation optics, left-handed
materials are precisely the complementary media of conventional dielectric
materials by mirror transformation\cite{10,11,12,13}. Under appropriate
design, a complementary object destructively interferes with an outside
object, and perfectly cancels the scattered waves from it, giving rise to
perfect cloak to the whole system\cite{14}. The idea is also successfully
applied to design perfect lens\cite{15}, the \textquotedblleft
superscatterer\textquotedblright\cite{16} and the \textquotedblleft
anticloak\textquotedblright\cite{17,18} as well. The concept of transformation
media can be extended to the scope of acoustic waves. A scheme to achieve
acoustic cloaking in two dimensions was proposed by Cummer and
Schurig\cite{19}. Subsequently, Huanyang Chen and C. T. Chan generalized the
idea to three dimensions by mapping the acoustic equation directly to the
conductivity equation\cite{20}. Recently, a technique to achieve broadband
acoustic cloak was proposed by J. B. Pendry and Jensen Li\cite{21}.

In this paper, we show that the concept of complementary media can be
generalized in acoustic waves. The acoustic complementary media possess
constitute parameters in opposite sigh of each other. Perfect acoustic
cloaking and perfect imaging with complementary media are demonstrated
numerically in two dimensions. The recovery of the acoustic wave field by
complementary media is interpreted in one dimensional multilayer system.

The acoustic wave equation takes a form as:%

\begin{equation}
\nabla\left[  \frac{1}{\rho\left(  x\right)  }\nabla p\left(  x\right)
\right]  =-\frac{\omega^{2}}{k\left(  x\right)  }p\left(  x\right)
\label{Eq-1}%
\end{equation}

where mass density $\rho\left(  x\right)  $and bulk modulus $k\left(
x\right)  $ are two constitutive parameters of the acoustic media at $x$
point. An alternative parameter equivalent to $k\left(  x\right)  $ is\ sound
velocity $c\left(  x\right)  $\ with $c^{2}\left(  x\right)  =k\left(
x\right)  /\rho\left(  x\right)  .$ We will use $\rho(x)\ $and $c(x)$\ instead
to characterize acoustic media later in full-wave simulations. Concerning the
acoustic complementary media, we start from a layered model by transforming a
single layer in acoustic wave space into two layers with mapping
functions\cite{15}. As shown in Fig.1, the original region in the $x^{\prime
}yz$\ Cartesian coordinate of acoustic virtual space with $x^{\prime}%
\in\left[  c,d\right]  $ now is transformed into two regions $x\in\left[
-a,0\right]  $ and $x\in\lbrack0,b]$\ in real space $xyz$ by mapping function
in Fig. 1(b). For simplicity, we adopt a linear form of mapping function with
$f(x):x=x^{\prime}$\ in the region of $x\in\left[  0,b\right]  $ , and
$g(x):x=-x^{\prime}$\ in the region of $x\in\left[  -a,0\right]  $. The
boundary continuum conditions of acoustic waves requires $f(x)$ and $g(x)$ to
satisfy with $g(0)=f(0)=c\ $and $g(-a)=f(b)=d$\ at the three interfaces
$x=-a$, $x=b$ and $x=0$. We obtain the constitutive parameters of the two
transformed layers 1 and 2 by the following equations:%

\begin{equation}%
\begin{array}
[c]{c}%
1/\rho_{1}=A_{g}\left[  1/\rho\right]  A_{g}^{T}/\det A_{g}\\
1/\rho_{2}=A_{f}\left[  1/\rho\right]  A_{f}^{T}/\det A_{f}%
\end{array}
\label{Eq-2}%
\end{equation}

\begin{equation}%
\begin{array}
[c]{c}%
k_{1}=k\det\left(  A_{g}\right) \\
k_{2}=k\det\left(  A_{f}\right)
\end{array}
\label{Eq-3}%
\end{equation}
Where$A_{g}=diag\left[  1/g^{\prime}(x),1,1\right]  =diag\left[
-1,1,1\right]  $,$A_{f}=diag\left[  1/f^{\prime}(x),1,1\right]  =diag\left[
1,1,1\right]  $ and \textquotedblleft$\det$\textquotedblright\ denotes the
determinant of matrix.

With the mapping functions satisfy to $g(x)=-f(x)$, layer 1 and layer 2 make
up of acoustic complementary media. And we have $-\rho_{1}=\rho_{2}=\rho$ and
$-k_{1}=k_{2}=k$\ for the mass density and bulk modulus of the two layers,
where $\rho$\ and $k$\ are the constitutive parameters of the original layer.
The transformation assigns exactly the same amplitude and phase to acoustic
pressure field at the boundary $x=-a$\ and at $x=b$ as if the two layers do
not exist.

We now numerically demonstrate how to realize two-dimensional acoustic cloak
with complementary media. Using a finite element solver (COMSOL Mutiphysics),
we calculate the acoustic scattering from a $0.1m\times0.2m$ sized rectangle
object incident from left to right by a monochromatic plane wave at frequency
$1.5kHz$. The host media is air. The mass density and the sound velocity of
the object are $\rho=2700kg/m^{3}$ and $c=5040m/s$\ respectively, Fig. 2(a)
presents the normalized pressure fields in colormap with respect to the
incidence. In Fig. 2(b), the host media $\left(  x<0\right)  $ and $\left(
x>0.2\right)  $ are air, the right region $\left(  0<x<0.2\right)  $ is
\textquotedblleft anti-air\textquotedblright\ with parameters $\rho
=-1.21kg/m^{3}$\ and $c=-343m/s$. The object, embedded in the air layer to be
cloaked, is exactly the same as that in Fig. 2(a). An \textquotedblleft
anti-object\textquotedblright, with the same size and shape of the object at
the left side, is embedded in the \textquotedblleft anti-air\textquotedblright%
\ layer. The object and the \textquotedblleft anti-object\textquotedblright%
\ are symmetrically positioned with respect to $\hat{y}$\ axis. The normalized
acoustic pressure fields are shown in Fig. 2. The regions where the normalized
amplitude of the pressure fields exceeds the bound of the color bar are marked
by white color. The absence of the scattered waves indicates invisibility of
the object and the anti-object.

To understand the perfect acoustic cloaking by complementary media in Fig.
2(b), we consider a one-dimensional layered system comprised of complementary
media. As shown in Fig. 3, the whole structure is the combination of two
trilayers $(+A+B+A)$ and $(-A-B-A)$. Layer $(+A)$ and layer $(-A)$ possess the
same thickness $d_{A}$ , layer $(+B)$\ and layer $(-B)$ possess the same
thickness $d_{B}$. The constitutive parameters of the layers and their
anti-layers satisfy with $\rho_{+A}=-\rho_{-A},c_{+A}=-c_{-A}$\ and $\rho
_{+B}=-\rho_{-B},c_{+B}=-c_{-B}\ $\ respectively. According to one-dimensional
transfer matrix method theory\cite{22} a centrosymmetric trilayer $\left(
+A+B+A\right)  $\ can be equivalent to one homogeneous layer with effective
characteristic impedance $R_{eff}\ $and phase thickness $\delta_{eff}$. The
scheme also applies for the trilayer $(-A-B-A)$, its characteristic impedance
$R_{eff}^{\prime}$\ and phase thickness $\delta_{eff}^{\prime}$\ satisfy with
$R_{eff}^{\prime}=R_{eff}$ and $\delta_{eff}^{\prime}=-\delta_{eff}$. The
characteristic matrix of an acoustic layer has a similar form of that of
transverse magnetic (TM) polarization in EM waves, so that we derive the
characteristic matrix of the whole structure as:%

\begin{equation}
M=M_{1}M_{2}=\left[
\begin{array}
[c]{cc}%
\cos\delta_{eff} & iR_{eff}\sin\delta_{eff}\\
\frac{i}{R_{eff}}\sin\delta_{eff} & \cos\delta_{eff}%
\end{array}
\right]  \left[
\begin{array}
[c]{cc}%
\cos\delta_{eff} & -iR_{eff}\sin\delta_{eff}\\
\frac{-i}{R_{eff}}\sin\delta_{eff} & \cos\delta_{eff}%
\end{array}
\right]  \label{Eq-4}%
\end{equation}

The fact that $M\ $regresses to an identity matrix means that the
complementary tri-layer as an anti-object rehabilitates the whole wave field
by canceling the scattering waves completely as if the whole system were
nothing but the host media.

We also use the concept of complementary media to demonstrate the imaging
effect in acoustic waves. As a specific example, in Fig. 4(a) we show the
pressure field distribution when a point source interacting with complementary
media and a cylinder object. A point source with frequency $3kHz$ is placed at
the position of $\left(  -0.2,0\right)  $\ the left boundary of a bilayer, in
which the left one is an air layer at $x\in\left[  -0.2,0\right]  $\ with
$\rho=1.21kg/m^{3}$, $c=343m/s$, and the right one is an \textquotedblleft
anti-air\textquotedblright\ layer at $x\in\left[  0,0.2\right]  $\ with
$\rho=-1.21kg/m^{3}$, $c=-343m/s$. A cylinder object with a radius of $0.07m$
and $\rho=27kg/m^{3}$, $c=400m/s$, is located at $\left(  -0.1,0.1\right)
$\ in the left air layer. As shown in Fig. 4(a), we can't get an image of the
source. If we set an \textquotedblleft anti-object\textquotedblright\ in the
\textquotedblleft anti-air\textquotedblright\ layer right side with same size
and shape of the cylinder object, and ensure them symmetrically positioned to
axis. Then we can get an image with good quality as illustrated in Fig. 4(b).
This implies that image effect can be achieve even an obstacle is placed in
front of the source.

In conclusion, we have generalized the concept of the complementary media in
acoustic waves. Using this concept, we show examples to cloak an object
outside and achieve perfect image in acoustic waves. Undoubtedly, this scheme
can be extend to three dimensions and not constrained by the shape of the
object. Recent research work on acoustic metamaterials have shown that bulk
modulus and mass density can be designed to vary in a wide range from negative
to positive\cite{23,24,25,26,27,28}, making transformation media attainable
practically in acoustic waves as well as complementary media. This work was
supported by the National 863 Program of China (Grant No.2006AA03Z407), NSFC
(GrantNo.10574099, No.60674778), NECT, STCSM and Shanghai Education and
Development Foundation (No. 06SG24).

\newpage%

\begin{figure}
[ptb]
\begin{center}

\includegraphics[
natheight=3.667in, natwidth=6.889in, height=3.667in, width=6.889in ]
{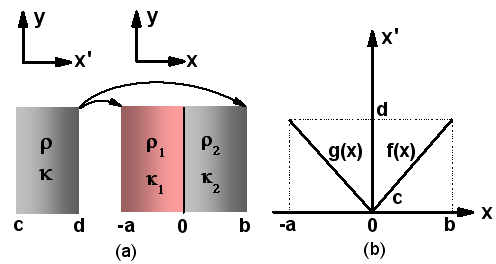}

\caption{(a) Coordinate transformation from a single slab to bilayers. (b) The
mapping function is $f(x):x=x^{\prime}$ in the region of $x\in\left[
0,b\right]  $ and $g(x):x=-x^{\prime}$\ in the region of \thinspace
$x\in\left[  -a,0\right]  $. The transformed double layers are complementary
each other when $f(x)=-g(x)$ and $a=b$.}%
\end{center}
\end{figure}

\begin{figure}
[ptb]
\begin{center}

\includegraphics[
natheight=6.833in, natwidth=4.931in, height=6.833in, width=4.931in ]
{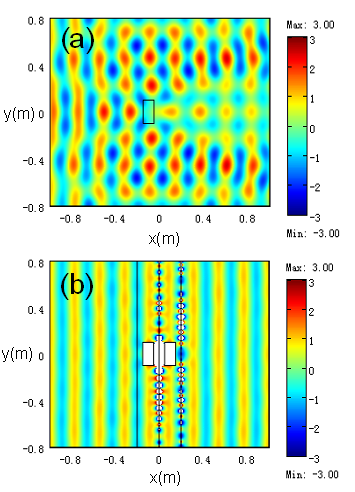}

\caption{Normalized acoustic pressure under an incident plane wave from left
to right. (a)An object in air with mass density $\rho=2700kg/m^{3}$\ and sound
velocity $c=5040m/s$. (b)An object in air is cloaked by complementary media.
An \textquotedblleft anti-object\textquotedblright\ with mass density
$\rho=-2700kg/m^{3}$\ and sound velocity $c=-5040m/s$ is embedded in the
\textquotedblleft anti-air\textquotedblright\ region $\left(  0<x<0.2\right)
$.}%
\end{center}
\end{figure}

\begin{figure}
[ptb]
\begin{center}

\includegraphics[
natheight=3.208in, natwidth=4.458in, height=3.208in, width=4.458in ]
{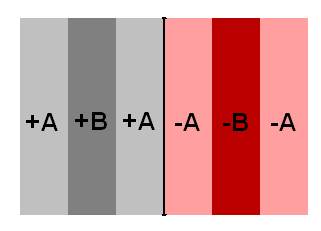}

\caption{One-dimensional acoustic layered system accounting for the perfect
cloak with complementary media. ($-$A) and ($-$B) are the complementary media
of A and B with $\rho_{+A}=-\rho_{-A},c_{+A}=-c_{-A}$\ and $\rho_{+B}%
=-\rho_{-B},c_{+B}=-c_{-B}.$}%
\end{center}
\end{figure}

\begin{figure}
[ptb]
\begin{center}

\includegraphics[
natheight=6.597in, natwidth=5.181in, height=6.597in, width=5.181in ]
{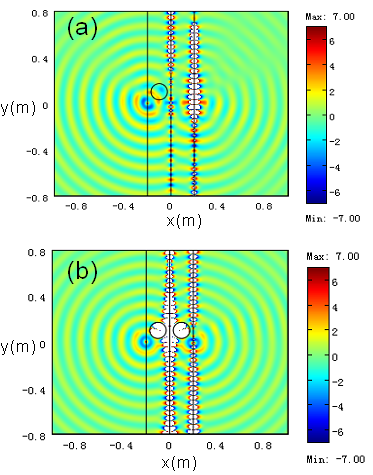}

\caption{(a) Pressure field distribution when a point source interacting with
complementary media and a cylinder object. The cylinder object with
$\rho=27kg/m^{3},c=400m/s$ is located at the position of $\left(
-0.1,0.1\right)  $. The left layer has $\rho=1.21kg/m^{3},c=343m/s$ located at
$x\in\left[  -0.2,0\right]  $. The right layer with $\rho=-1.21kg/m^{3}%
,c=-343m/s$ is located at $x\in\left[  0,0.2\right]  $, The host media
$\left(  x<0\right)  $ and $\left(  x>0.2\right)  $ are air. (b) Pressure
field distribution when an \textquotedblleft anti-object\textquotedblright%
\ with $\rho=-27kg/m^{3},c=-400m/s$ is embedded in the right layer at the
position of $\left(  0.1,0.1\right)  $.}%
\end{center}
\end{figure}

\end{document}